\input epsf
\documentclass[draft]{aa}
\begin{document}

%\thesaurus{20(08.05.2, 13.09.6)} 

\title
{A representative sample of Be stars III:  \newline 
$H$ band spectroscopy}

\author{I.~A.~Steele\inst{1} 
\and J.~S.~Clark\inst{2,3}
}
\institute{Astrophysics Research Institute, Liverpool John Moores
University, Liverpool, CH41 1LD, U.K.
\and Astronomy Centre, CPES, University of Sussex, Brighton, BN1 9QH, 
U.K.
\and Dept. of Physics \& Astronomy, University College London, London, 
WC1E 6BT, U.K.}

\offprints{I. A. Steele}
\mail{ias@astro.livjm.ac.uk}

\date{Received    / Accepted     }
\titlerunning{A representative sample of Be stars III}
\authorrunning{Steele \& Clark}

\abstract{
We present 
$H$ band (1.53$\mu$m-1.69$\mu$m) spectra of 57 isolated Be stars 
of spectral types O9-B9 and luminosity classes III,IV \& V. The H\,{\sc i} 
Brackett (n--4) series is seen in emission from Br-11--18, and  Fe\,{\sc ii}
emission is also apparent for a subset of those stars with H\,{\sc i} emission.
No emission from species with a higher excitation temperature, such as 
He\,{\sc ii} or C\,{\sc iii} is seen, and no  forbidden line emission is 
present. A subset of 12 stars show no evidence for emission from any species; 
these stars appear indistinguishable from normal B stars of a comparable 
spectral type. 
In general the 
line ratios constructed from the transitions in the range Br-11--18
do not fit case B recombination theory particularly well.  
Strong correlations between the line ratios with Br-$\gamma$ 
and spectral type are found. 
These results most likely represent systematic variations in the temperature
and ionization of the circumstellar disc with spectral type.
Weak correlations between the line widths and projected rotational velocity 
of the stars are
observed; however no systematic trend for increasing line width through the 
Brackett series is observed.
}

\maketitle

\section{Introduction}

Be stars are defined as hot, non-supergiant stars that show, or have
shown at some stage, emission lines in their spectrum.   About 
20\% of B stars are Be stars hence they are an
important part of the hot star population.   Understanding them
is crucial to our obtaining a complete picture of hot-star winds.
They are rapid rotators (rotating at a mean velocity of 70\% of their break-up
speeds, Porter 1996) and often
have complex and variable emission line profiles that at sufficient
resolution are 
double peaked.  In addition they show an optical and infrared
continuum excess.  In spite of many attempts, no complete explanation for the
Be phenomenon has yet been found.  
 
Observations have produced an empirical
description of Be star circumstellar environments which is now
generally accepted: a dense ``disc''
exists in the equatorial plane (probably rotating in a Keplerian
fashion), whilst over the polar regions there is a fast wind
(velocities up to $\sim 1000-2000$km s$^{-1}$).
The disc is ionized,
and it is recombination in the disc that gives the emission lines.
The double peaked structure of these lines is then due to the
velocity structure and to self-absorption in the disc.  
Gehrz et al. (1974)
showed that the optical and infrared excess in the systems could be
explained by disc free-free emission.   Recent radio and optical  
interferometric data has confirmed the existence of the
disc (Dougherty \& Taylor 1992, Quirrenbach et al. 1994, Stee et al. 1995). 

Studies of Be stars have traditionally concentrated on
optical--near IR photometry and optical spectroscopy. However, the
optical H\,{\sc i} recombination lines do not
function as good diagnostics of the inner regions of the circumstellar
disc. Since it is likely that studying the inner regions of the disc
will result in important constraints to the physical processes giving rise
to the different wind regimes the need for observations of this region
are pressing. Near--IR spectroscopy provides one tool to probe the
innermost regions of the circumstellar disc. We have therefore obtained
H and K band spectra of a sample of some $\sim$60 Be stars, from
B0--B9 to study this region of the circumstellar envelope for the
first time. 

This paper is the third of a series on the optical and
near IR spectral properties of a representative sample of 58 Be stars.
In Steele et al. (1999; Paper I) we discuss the the
basic properties of our sample, such as spectral type, luminosity
class and projected rotational velocity, and determine that no
significant selection effects bias our sample. 
In Clark \& Steele (2000; Paper II) we
present the K band spectra of the sample and relate these to the
underlying properties of the stars.  In this paper 
we present $H$ band (1.53--1.69 $\mu$m) spectra 
of 55 of the stars from Paper I, plus three
additional objects (see Table 7).

\begin{table*}
\caption{K Band Spectral Groups and their corresponding
optical spectral types (see Paper II for details)}
\begin{center}
\begin{tabular}{lll}
\hline
Group & K Band appearance & Spectral Type \\
\hline
1 & Br $\gamma$ emission, He {\sc i} features & O9e -- B3e \\
2 & Br $\gamma$ absorption, He {\sc i} features & O9 -- B3 \\
3 & Br $\gamma$ + Mg {\sc ii} emission, no He {\sc i} features & B2e -- B4e \\
4 & Br $\gamma$ absorption, no He {\sc i} or Mg {\sc ii} features & B4 -- B9 \\
5 & Br $\gamma$ emission,  no He {\sc i} or Mg {\sc ii} features & 
B4e -- B9e \\
\hline
\end{tabular}
\end{center}
\end{table*}

\section{Observations \& Data Reduction}

The sample of target objects contains objects from
O9 to B8.5 and of luminosity classes III (giants) to V (dwarfs), 
as well as three shell stars.  
The sample was selected in an attempt to contain several objects that were 
typical of each spectral
and luminosity class in the above range; it therefore does {\em not}
reflect the spectral and luminosity class space distribution of Be stars,
but only the average properties of each subclass in temperature and
luminosity. A spectral type and measure 
of $v \sin i$ was derived for each object in the sample and were presented
in Paper I (only the spectral classifications are repeated here
 for sake of brevity). 
The distribution of $v \sin i$ within each temperature and
luminosity class was  carefully investigated and the conclusion
drawn that there were no significant selection effects biasing the
average properties of the objects (see Paper I for details).

\def\epsfsize#1#2{0.47#1}
\begin{figure}
\setlength{\unitlength}{1.0in}
\centering
\begin{picture}(3.3,3.8)(0,0)
\put(-0.4,-0.7){\epsfbox[0 0 2 2]{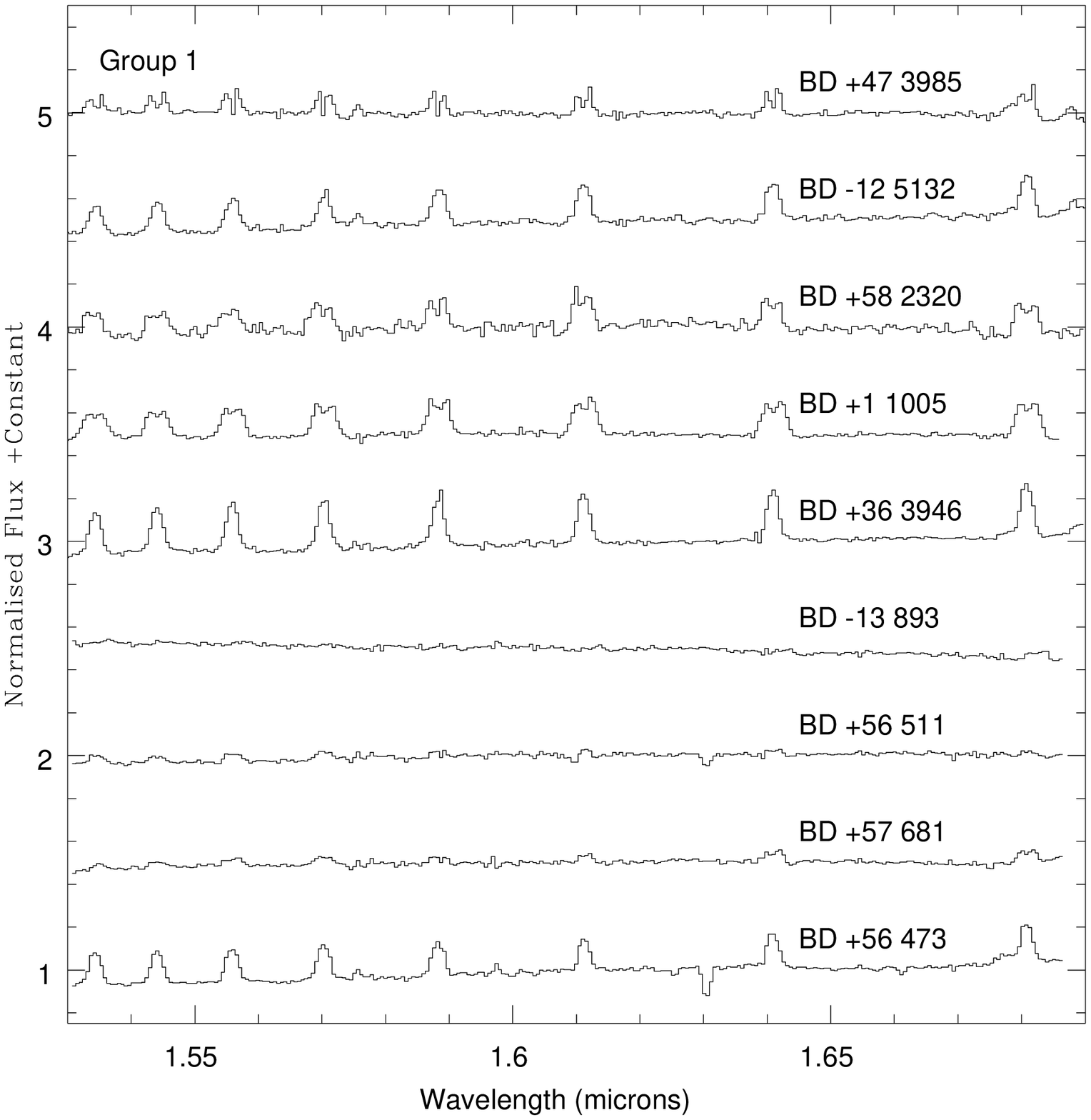}}
\end{picture}
\caption{ H band spectra for Group 1 objects (I).}
\end{figure}

\def\epsfsize#1#2{0.47#1}
\begin{figure}
\setlength{\unitlength}{1.0in}
\centering
\begin{picture}(3.3,3.8)(0,0)
\put(-0.4,-0.7){\epsfbox[0 0 2 2]{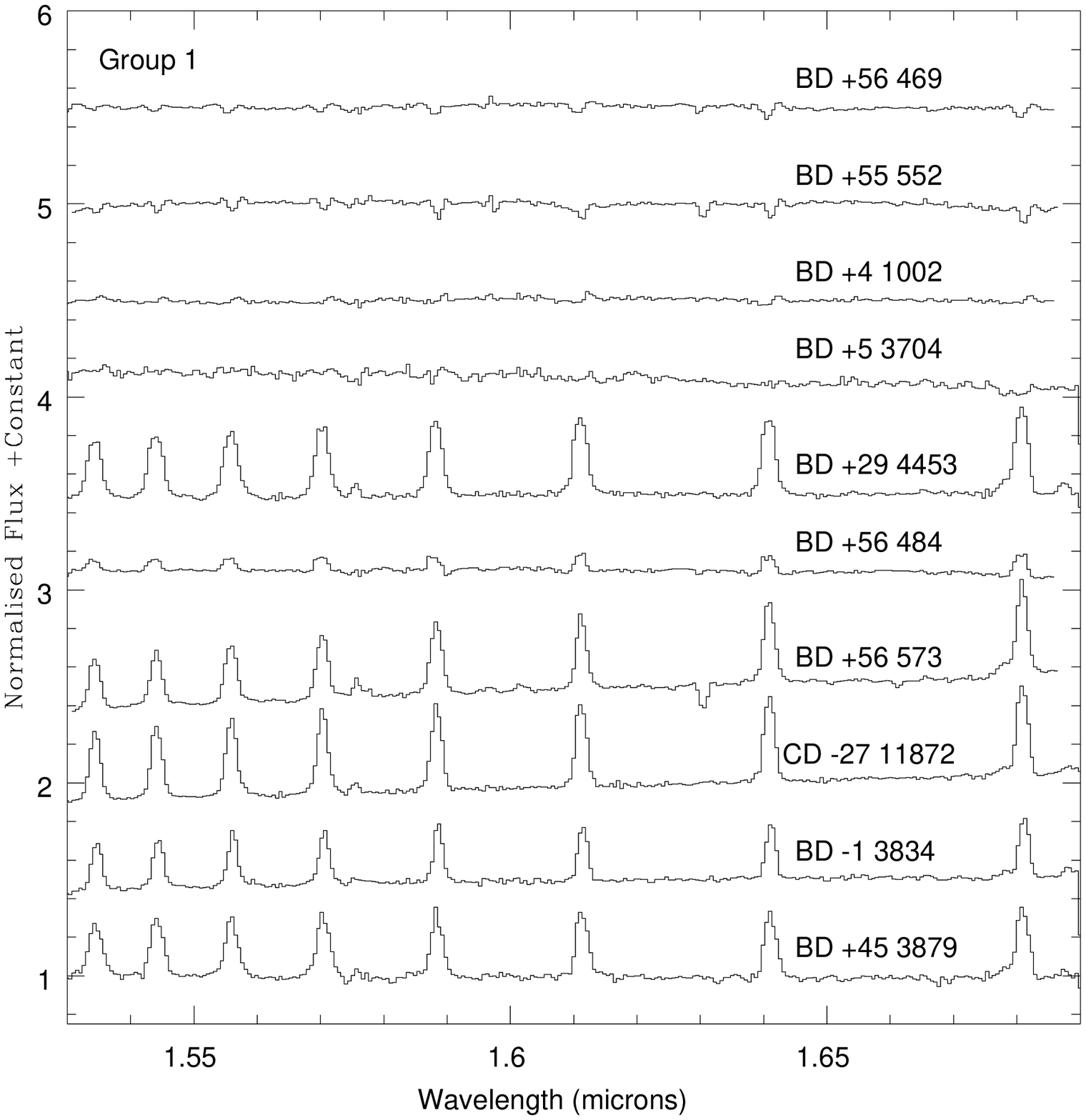}}
\end{picture}
\caption{ H band spectra for Group 1 objects (II).}
\end{figure}

\def\epsfsize#1#2{0.47#1}
\begin{figure}
\setlength{\unitlength}{1.0in}
\centering
\begin{picture}(3.3,3.8)(0,0)
\put(-0.4,-0.7){\epsfbox[0 0 2 2]{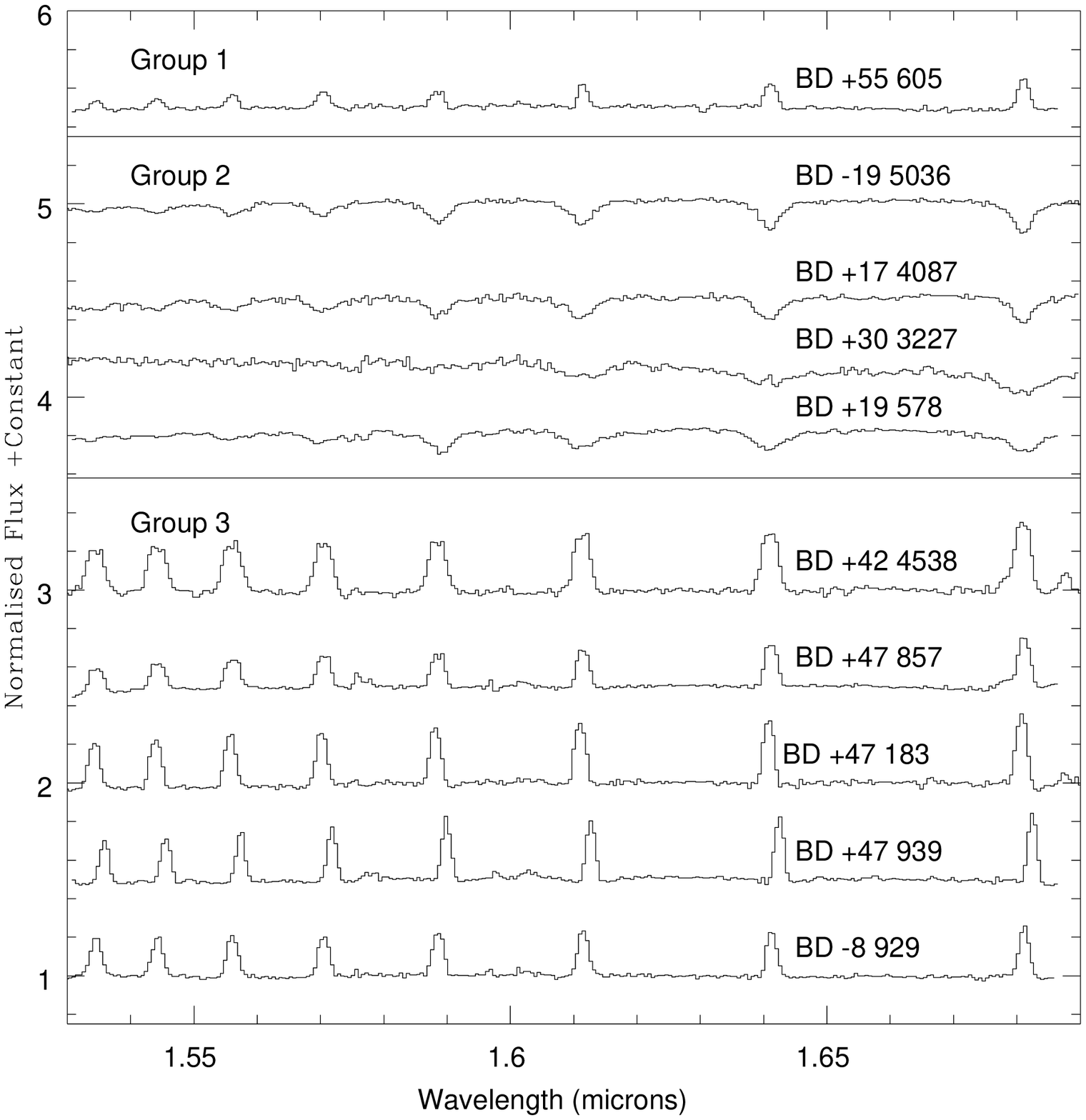}}
\end{picture}
\caption{ H band spectra for Group 1 (top panel) objects (III), 
Group 2 (middle panel) objects and 
Group 3 objects (bottom panel).}
\end{figure}

\def\epsfsize#1#2{0.47#1}
\begin{figure}
\setlength{\unitlength}{1.0in}
\centering
\begin{picture}(3.3,3.8)(0,0)
\put(-0.4,-0.7){\epsfbox[0 0 2 2]{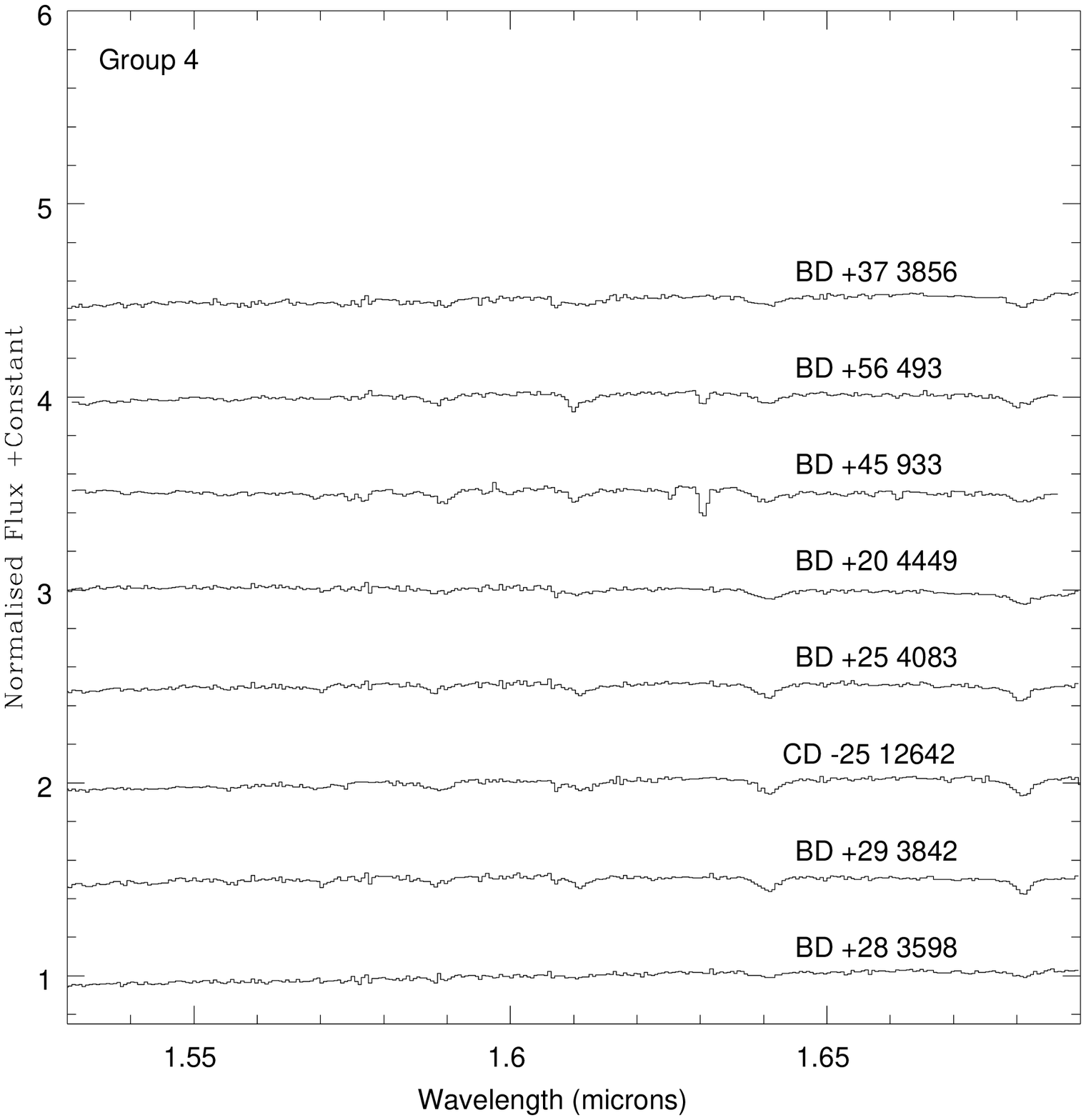}}
\end{picture}
\caption{ H band spectra for Group 4 objects.}
\end{figure}

\def\epsfsize#1#2{0.47#1}
\begin{figure}
\setlength{\unitlength}{1.0in}
\centering
\begin{picture}(3.3,3.8)(0,0)
\put(-0.4,-0.7){\epsfbox[0 0 2 2]{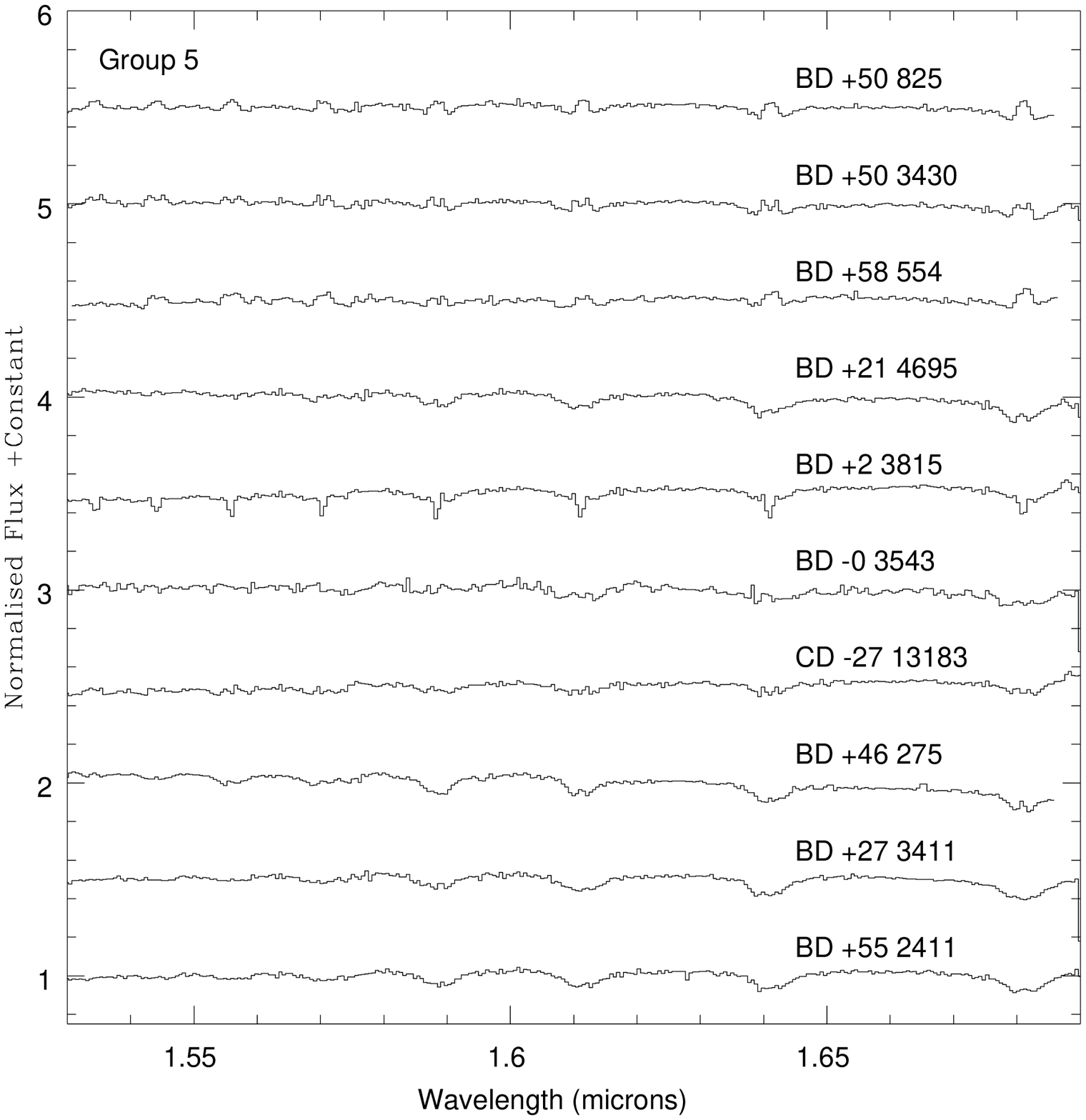}}
\end{picture}
\caption{ H band spectra for Group 5 objects (I).}
\end{figure}

\def\epsfsize#1#2{0.47#1}
\begin{figure}
\setlength{\unitlength}{1.0in}
\centering
\begin{picture}(3.3,3.8)(0,0)
\put(-0.4,-0.7){\epsfbox[0 0 2 2]{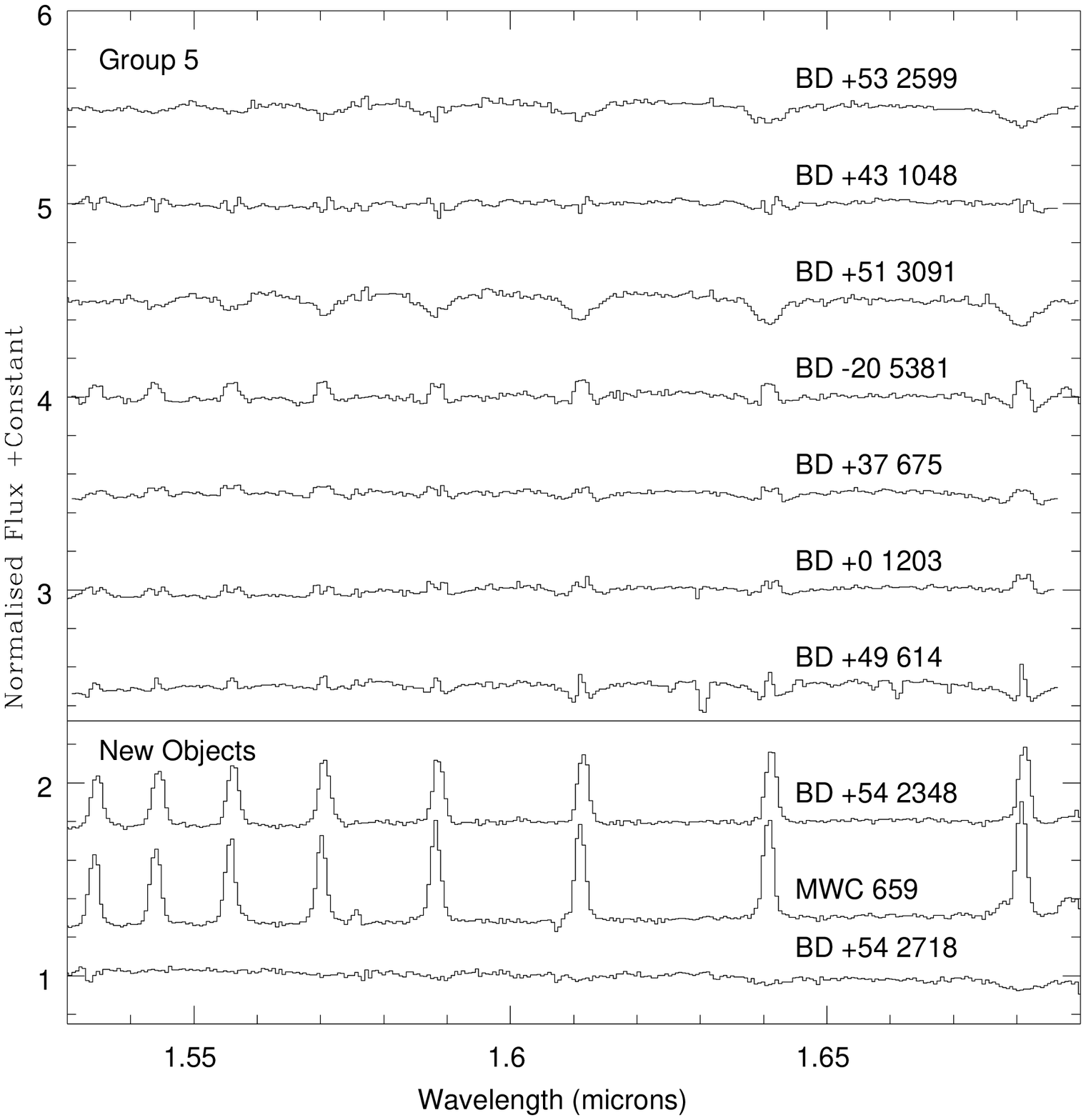}}
\end{picture}
\caption{ H band spectra for Group 5 (upper panel) objects (II) 
and new objects (lower panel).}
\end{figure}

The observations were carried out on the United Kingdom Infrared
Telescope (UKIRT) on 1996 June 29 and October 1--2 
(see Tables 2--7), using the Cooled Grating Spectrometer
(CGS4).   The observations were made 
using the short focal length camera 
plus the 150 line/mm grating, giving an ideal coverage
from 1.53 to 1.71$\mu$m with a velocity resolution of $\sim 100$ km/s.
Unfortunately, due to problems with the slit rotation mechanism,
the wavelength coverage was slightly curtailed, 
leading to a final wavelength coverage of 1.53 to 1.69 $\mu$m, 
excluding the He\,{\sc i} 1.7004$\mu$m transition. 
Data reduction was carried out in a similar manner as described in Paper II
with
correction for telluric features by 
ratioing observed G type standards with the solar spectrum to
remove features within their photosphere, and then ratioing the
object spectrum with the modified standard spectrum.

\begin{table*}
\begin{center}

\caption{Summary of measured H\,{\sc i} Brackett series, He\,{\sc i} and 
Fe\,{\sc ii} equivalent widths in {\AA}
for Group 1 stars.  Note that no correction is made in these tables for
the underlying photospheric absorption line, and that 
a {\em positive} EW corresponds to {\em emission}.
{\em e} indicates emission in the relevant transition that is to
weak to measure (typically $<0.2${\AA}), the
addition of a question mark represents an uncertainty as to its presence.
A dash indicates that the feature was not within the wavelength
range of that particular spectrum.
The estimated error in the equivalent widths is 20 \%.
The spectral type of each star as
determined in Paper I is given in column 2 (where an italicized type indicates
that only a historical classification is available), and the 
observation date during 1996 is indicated in column 3.}

\begin{tabular}{llcccccccccccc}
\hline
Object  & Spec.&Obs.&Br18 & Br17 & Br16 & Br15  & Fe\,{\sc ii} & Br14 & Br13 &
Br12 & Fe\,{\sc ii} & Br11 & Fe\,{\sc ii} \\
 &Type&Date&&      &      &       & 1.576        &      &      &   & 1.679 & & 1.687 \\ 
\hline
CD -27 11872& B0.5V-III    & 29/6 & 8.3 &  9.2 & 9.8 & 10.0&0.8 & 10.0& 9.7& 9.6 & e&9.6 & e\\
BD -13 893  & {\em B1-3V}  & 1/10 & 0.0 & 0.0 & 0.0 & 0.0 & 0.0 & 0.0& 0.0& 0.2 & 0.0& 1.4& -\\
BD -12 5132 & BN0.2III     & 29/6 & 2.8 & 3.1 & 3.5 & 3.9  &1.0 &4.7  &4.6 &4.1 & e&3.9 & e\\
BD -1  3834 & B2IV         & 29/6 &  4.6  &5.3  &5.6  & 6.2& 0.8 & 6.3& 6.2& 5.7 & e & 5.6 & e\\
BD +1  1005 & {\em B1-3V}  & 2/10 & 4.9&5.0& 5.5& 6.3 & 0.0  &6.1 &5.9 &5.6 &0.0 &5.7 & -\\
BD +4  1002 & {\em B2-3III}& 2/10 & 0.2  &0.2 & 0.3  &0.2  & 0.0 & 0.3  &0.7 &0.1 & 0.0 &0.1 & -\\
BD +5 3704  & B2.5V        & 29/6 & 0.0  & 0.0&0.0&0.1&0.0&0.2&0.0&0.0 &0.0 & -0.2& 0.0\\
BD +29 4453 & B1.5V        & 29/6 & 8.7 &9.7  &9.9  &10.2& 0.9 & 10.6& 11.5 &10.5 & e& 11.7& e\\
BD +36 3946 & B1V          & 29/6 & 4.6    &  5.5& 5.5  &5.6  & 0.4& 6.0& 5.7& 5.5 & e & 5.4 & e\\
BD +45 3879 & B1.5V        & 29/6 & 7.8 & 7.8 & 7.8  & 8.5 &0.7  &7.7 &8.4 & 7.8&e & 9.3& e?\\
BD +47 3985 & B1-2sh       & 29/6 & 1.8 &2.1  &2.4   &1.9 &0.2  &1.8 &2.1 & 2.4 &e & 4.1&e \\
BD +55 605  & B1V          & 1/10 & 0.7& 0.9 &1.4 & 2.1  & 0.0 &2.1 &2.4 &2.9 &0.0 &3.2 & -\\
BD +55 552  & B4V          & 1/10 & 0.1 & -0.4 & -0.1& -1.2  &0.0 &-1.0 &-1.1 & -3.0 &0.0&-0.3& -\\
BD +56 469  & {\em B0-2III}& 1/10 & -0.4 & -0.4 & -0.3 &-0.3  & 0.0& -0.3 &-0.3  &-0.6&0.0&-0.6&-\\
BD +56 473  & B1V-III      & 2/10 & 3.4& 3.5 &3.7  &3.9 &0.5 & 4.0 &3.9 & 3.8& e& 4.5& - \\
BD +56 478  & B1.5V        & 1/10 & 0.9  &1.2 & 1.4 &1.5  & 0.0 & 1.5& 1.4& 1.9& 0.0 &2.1 & - \\
BD +56 511  & B1III        & 1/10 & 0.5  &0.4 & 0.7  &1.0 & 0.0 & 0.1 & 0.3& 0.7& 0.0& 0.3 & -\\
BD +56 573  & B1.5V        & 1/10 &  5.9& 7.2& 7.8   & 8.6  &2.1 & 9.8& 8.8& 9.7&e & 10.3& -\\
BD +57 681  & B0.5V        & 1/10 & 0.6  &0.4  &0.9 &1.2 & 0.0 &0.9 & 1.0& 1.6 &0.0 &2.0 & -\\
BD +58 2320 & B2V          & 29/6 & 2.9& 3.6  &4.1  &4.0   &0.0 &3.2 & 4.6&3.8 &0.0 &5.9 & 0.0\\
\hline
\end{tabular}
\end{center}
\end{table*}

\begin{table*}
\begin{center}
\caption{Summary of H\, {\sc i} Brackett series for Group 2 stars.
See Table 2 for explanation}
\begin{tabular}{llcccccccccccc}
\hline
Object  & Spec. &Obs. &Br18 & Br17 & Br16 & Br15  & Fe\,{\sc ii} & Br14 & Br13 &
Br12 & Fe\,{\sc ii} & Br11 & Fe\,{\sc ii} \\
        &Type &Date &&      &      &       & 1.576        &      &      &   & 1.679 & & 1.687 \\ 
\hline
CD -25 12642 &B0.7III  &29/6 & -0.3&-0.5 &-0.4 & -0.8&0.0& -1.7& -2.9& -3.0 &0.0& -4.0& 0.0\\
BD +20 4449  &B0III    &29/6& 0.0& 0.0& 0.0& -0.5&0.0 &-1.0 &-1.7 &-2.0 & 0.0& -3.0& 0.0\\
BD +25 4083  &B0.7III  &29/6& -0.4& -0.2& -0.1& -0.7&0.0 &-1.5 &-2.1 & -2.5& 0.0& -3.2&0.0 \\
BD +28 3598  &O9II     &29/6& 0.0& 0.0& 0.0& 0.0&0.0 &-0.3 & 0.3& -0.7& 0.0&-1.5& 0.0\\
BD +29 3842  &B1II     &29/6& -0.3& -0.3& -0.3& -0.8&0.0 &-1.4 &-1.6 &-2.3 &0.0 & -2.3& 0.0\\
BD +37 3856  &B0.5V    &29/6&0.0 & 0.0& 0.0& -0.5&0.0 &-0.8 &-1.7 &-2.3 & 0.0& -2.8& 0.0\\
BD +45 933   &B1.5V    &1/10&  -0.5& -0.9& -1.1& -1.6&0.0 & -1.8&-1.3 & -2.0& 0.0& -1.9& - \\
BD +56 493   &B1V      &1/10& -0.3& -0.1& -0.2& -0.7&0.0 &-1.3 & -2.3& -1.7& 0.0& -2.9& - \\
\hline
\end{tabular}
\end{center}
\end{table*}

\begin{table*}
\begin{center}
\caption{Summary of H\, {\sc i} Brackett series for Group 3 stars.
See Table 2 for explanation}
\begin{tabular}{llcccccccccccc}
\hline
Object  & Spec. &Obs. &Br18 & Br17 & Br16 & Br15  & Fe\,{\sc ii} & Br14 & Br13 &
Br12 & Fe\,{\sc ii} & Br11 & Fe\,{\sc ii} \\
 &Type  &Date &&      &      &       & 1.576        &      &      &   & 1.679 & & 1.687 \\ 
\hline
BD -8 929    &{\em B2V} &2/10& 4.6&4.5 &4.5 & 4.6&0.0 &5.0 &5.2 & 4.6& 0.0& 4.9& - \\
BD +42 4538  &B2.5      &29/6& 8.2& 8.7&9.2 &8.4 &0.0 & 9.0& 9.3&9.2 & 0.0& 10.6& 0.0\\
BD +47 183   &B2.5V     &29/6& 5.6& 5.6&5.7 & 5.8&0.3 &5.6 & 6.4& 6.3& e& 7.0& e \\
BD +47 857   &B4IV      &1/10& 2.4& 3.5&4.3 &4.4 &1.5 &4.4 & 4.5& 5.3& e& 5.7& -\\
BD +47 939   &B2.5V     &1/10& 3.6& 4.1& 4.3& 4.6&0.9 &5.4 &5.2 & 5.2& e& 6.4& -\\
\hline
\end{tabular}
\end{center}
\end{table*}

\begin{table*}
\caption{Summary of H\, {\sc i} Brackett series for Group 4 stars.
See Table 2 for explanation}
\begin{center}
\begin{tabular}{llcccccccccccc}
\hline
Object  & Spec. &Obs. &Br18 & Br17 & Br16 & Br15  & Fe\,{\sc ii} & Br14 & Br13 &Br12 & Fe\,{\sc ii} & Br11 & Fe\,{\sc ii} \\
  &Type&Date &&      &      &       & 1.576        &      &      &   & 1.679 & & 1.687 \\ 
\hline
BD -19 5036  &B4III     &29/6& -0.8& -1.4& -2.0& -2.5&0.0 & -4.8& -5.4&-6.3 & 0.0 & -7.5 &0.0\\
BD +17 4087  &B6III-V   &29/6& -1.2& -1.4& -2.1& -2.9&0.0 &-3.5 &-5.1 &-5.2 & 0.0 & -5.7 &0.0\\
BD +19 578   &B8V       &1/10& -0.3& -0.4& -0.4& -1.2&0.0 &-2.1 & -4.1& -5.3& 0.0 & -5.9& - \\
BD +30 3227  &B4V       &29/6& 0.0& 0.0& 0.0& -1.0& 0.0 &-3.4 &-5 & -5.7& 0.0 & -7.5&0.0\\
\hline
\end{tabular}
\end{center}
\end{table*}

\begin{table*}
\caption{Summary of H\, {\sc i} Brackett series for Group 5 stars.
See Table 2 for explanation.}
\begin{center} 
\begin{tabular}{llcccccccccccc}
\hline
Object  & Spec. &Obs. &Br18 & Br17 & Br16 & Br15  & Fe\,{\sc ii} & Br14 & Br13 &
Br12 & Fe\,{\sc ii} & Br11 & Fe\,{\sc ii} \\
 &Type& Date&  &      &      &       & 1.576        &      &      &   & 1.679 & & 1.687 \\ 
\hline
CD -27 13183 &B7V        &29/6& 0.0& 0.0&-1.0 &-1.4 &0.0 & -2.0 &-3.1 & -3.2& 0.0 & -4.80&0.0 \\
BD -20 5381  &B5V        &29/6& 1.0& 1.0& 1.7& 1.1 &0.0 & 0.7& 1.4& 0.7 & 0.0 & 0.8 & - \\
BD -0 3543   &B7V        &29/6& 0.0& 0.0& 0.0& -1.2&0.0 & -2.2& -2.4& -2.7& 0.0 & -3.3 & 0.0\\
BD +0 1203   &{\em B5III}&29/6& 1.1& 0.7& 0.7& 0.8&0.0 &0.9 &0.8 & 1.2& 0.0 & 2.0& 0.0 \\
BD +2 3815   &B7-8sh     &29/6& -0.5& -0.8& -1.2& -2.0&0.0 & -3.7& -3.5&-4.1 & 0.0 & -4.6&0.0 \\
BD +21 4695  &B6III-V    &29/6& 0.0&-0.3 &-0.8 &-1.3 &0.0 &-2.8 & -3.4& -5.4& 0.0& -6.4& 0.0 \\
BD +27 3411  &B8V        &29/6&0.0 &-0.4 &-0.7 &-1.1 &0.0 &-2.4 & -4.5& -5.6& 0.0& -6.5& 0.0 \\
BD +37 675   &B7V        &1/10& 1.2& 0.6&1.0 & 0.0 &e & -0.2& -1.7& -2.0& e & -1.5& - \\
BD +43 1048  &B6IIIsh    &1/10& 0.2& 0.2& 0.2& 0.3&0.0 &0.5 & 0.7& 1.4& 0.0&1.2 & - \\
BD +46 275   &B5III      &2/10& -0.1& -0.9& -2.2& -2.4&0.0 & -4.9& -4.3& -5.2& 0.0& -6.0& - \\
BD +49 614   &B5III      &1/10& 0.2& 0.3& 0.6& 0.5&0.0 & -1.6& -3.3 & -2.7& 0.0& -2.7& - \\
BD +50 825   &B7V        &2/10& 0.3 & 0.3& 0.2& -1.0&0.0 & -0.2& -1.7&-2.0& 0.0& -1.5& - \\
BD +50 3430  &B8V        &29/6& 0.5& 0.5& 0.6& -1.7&0.0 & -1.7&-1.9 &-2.6 & 0.0& -4.5& 0.0 \\
BD +51 3091  &B7III      &29/6& -1.0& -1.8& -2.4& -3.8&0.0 &-4.3 & -6.2& -6.8& 0.0& -6.7& 0.0 \\
BD +53 2599  &B8V        &29/6& -1.1& -1.9& -1.9& -2.7&0.0 & -2.4& -3.7& -5.1& 0.0& -7.5& 0.0 \\
BD +55 2411 & B8.5V      &29/6&-0.4& -0.6& -1.2& -2.1&0.0 & -4.0& -5.0 & -6.0 &0.0 &-7.5& 0.0 \\
BD +58 554   &B7V        &1/10& 0.0 &0.8 &1.2 & 0.7&0.0 & 0.2& 0.7& 0.6& 0.0 & 0.3& - \\
\hline
\end{tabular}
\end{center}
\end{table*}

\begin{table*}
\caption{Summary of H\, {\sc i} Brackett series for new stars. See Table 2
for explanation.}
\begin{center}
\begin{tabular}{llcccccccccccc}
\hline
Object  & Spec. & Obs. & Br18 & Br17 & Br16 & Br15  & Fe\,{\sc ii} & Br14 
& Br13 & Br12 & Fe\,{\sc ii} & Br11 & Fe\,{\sc ii} \\
        &  Type &Date& &      &      &       & 1.576        &      &      &   & 1.679 & & 1.687 \\ 
\hline
BD +54 2348  &{\em  B2V} &29/6&6.2& 7.0& 7.3& 7.9&0&7.6 &7.7 &7.9& 0.0 &8.7& 0.0 \\
BD +54 2718&{\em B2 III} &29/6&-0.5&-0.4& 0.0& 0.0&-0.7 &-1.0& -0.9& -1.4& 0.0 &-3.0&0.0\\
MWC 659 & {\em BOIIIpe}  &29/6& 7.8& 8.9& 9.8& 9.4& 0.8&10.7&10.3&10.9&e& 11.3 & e\\
\hline
\end{tabular}
\end{center}
\end{table*}

\section{Discussion}

\subsection{Line Identification}

The primary goal of these observations was to observe the H\,{\sc i}
Brackett recombination series.  The wavelength range chosen encompasses
Br-18--11, as well as Fe\,{\sc ii} lines at 1.534, 1.600 and 1.620$\mu$m
(amongst others).  Various He\,{\sc ii} transitions are also 
found within this wavelength region, however given their absence in the 
K band spectra
presented in Paper II, we do not expect to see these in emission. The
spectra are presented in Figs. 1--6, and are  presented 
in the same Groups we defined in Paper II,  which were based on
their K band spectral morphology (see Table 1). 
This approach has been adopted, rather than
grouping the spectra on the basis of their optically derived spectral
classifications since one aim of this work was to define a
classification scheme based on near--IR spectroscopy alone, for use in
the classification of heavily obscured stars, such as
those often found in High Mass X--ray binary systems.

Unsurprisingly, based on the features present in the K band spectra, 
the obvious lines seen in emission
are Br-18--11, and Fe\, {\sc ii} 1.576$\mu$m and 1.687$\mu$m. 
A further feature 
in the blue wing of Br-11 ($\lambda \sim$1.678$\mu$m) 
is also seen in a subset of spectra, the identity of which is discussed 
further below. 
No emission from species  of higher excitation (such as 
He\,{\sc ii}, N\,{\sc iii} or C\,{\sc iii}) was observed.
The presence and equivalent  widths (EW) 
of these features are summarised in Tables 2--7.  Equivalent widths 
and FWHM were measured using the {\sc abline} routine of the
{\sc figaro} software package.  This measures
EW by integrating the line flux relative to an interpolated continuum
produced by polynomial fitting in the vicinity of the line - it therefore
makes no assumption about line shape.  FWHM are measured by looking at
the line width which encompasses 68\% of the line flux and multiplying
by 1.18.  For a line of Gaussian profile this gives exactly the
FWHM, and for any reasonably centrally concentrated profile (as ours are)
it gives a result very close to the FWHM that is more robust than
simple Gaussian fitting. 

Brackett series emission is seen in all stars of Groups 1, 3 and 
5 (which all possess Br$\gamma$ emission), with the possible exception 
of BD +51 3091, a B7 III star (Fig. 6). Of Groups 2 and 4, which were 
defined on the basis of the absence of Br$\gamma$ emission in the  K band
spectra, only  BD +30 3227 appears to show evidence for partial infilling of
 Br-11--14 (Fig. 3), while the remainder appear to possess pure absorption   
spectra. 

Fe\, {\sc ii} 1.576$\mu$m 
emission is seen in a total of 13 stars, compared to a total of 19 stars 
showing Fe\,{\sc ii} 2.089$\mu$m emission.
Of the 33 stars where the wavelength coverage encompassed Fe\,{\sc ii} 
1.688$\mu$m, the line was in emission in 6 stars, 5 of which
are Group 1 objects, the final star, MWC 659, being one of the 3 objects 
without corresponding K band observations (see Tables 2--7). 
We note that all stars with 
Fe\, {\sc ii} 1.576$\mu$m emission also show  Fe\,{\sc ii} 1.688$\mu$m 
emission.   
The correlation between the presence of Fe\, {\sc ii} emission 
in the H and K bands is weaker, with a total of 8 stars showing Fe\, {\sc ii}
emission in either the H or K band, but not in the other. However, as 
with the K band spectra, Fe\, {\sc ii} emission only occurs in the H band
 spectra of those stars with strong Brackett series emission;
EW$_{Br-11} >$3.5{\AA}\footnote{Note that in this
paper we will employ the convention that {\em positive} equivalent widths 
indicate {\em emission} features.} (EW$_{Br{\gamma}} >$8{\AA}).

Two possible identifications exist for the emission feature at 
$\sim$1.678$\mu$m; Fe\,{\sc ii} 1.679$\mu$m and [Fe\,{\sc ii}] 1.678$\mu$m. 
Of these, we favour Fe\,{\sc ii} 1.679$\mu$m since the feature is only seen in
stars that show emission in the other two Fe\,{\sc ii}  lines, and no emission
is seen in the (albeit weaker) [Fe\,{\sc ii}] transitions at 1.534$\mu$m 
and 1.644$\mu$m (Hamann \& Persson 1989).

Overall, as regards spectral classification using the $H$ band
spectra of Be stars, it is apparent that with the absence of many pure 
photospheric features that are uncontaminated by disc emission it is only 
possible to perform a very general spectral classification into ``early'' 
(B0e--B4e) and ``late'' (B5e--B9e) spectral types.
The problem of the lack of photospheric features 
uncontaminated by emission from the circumstellar disc is compounded by
the (expected) lack of emission from species with a wide range of excitation
energies (although we speculate that the presence of He\, {\sc i} 1.700$\mu$m 
emission may function as an additional, valuable diagnostic of early 
spectral types).

\def\epsfsize#1#2{0.45#1}
\begin{figure}
\setlength{\unitlength}{1.0in}
\centering
\begin{picture}(3.3,3.3)(0,0)
\put(-0.2,-0.7){\epsfbox[0 0 2 2]{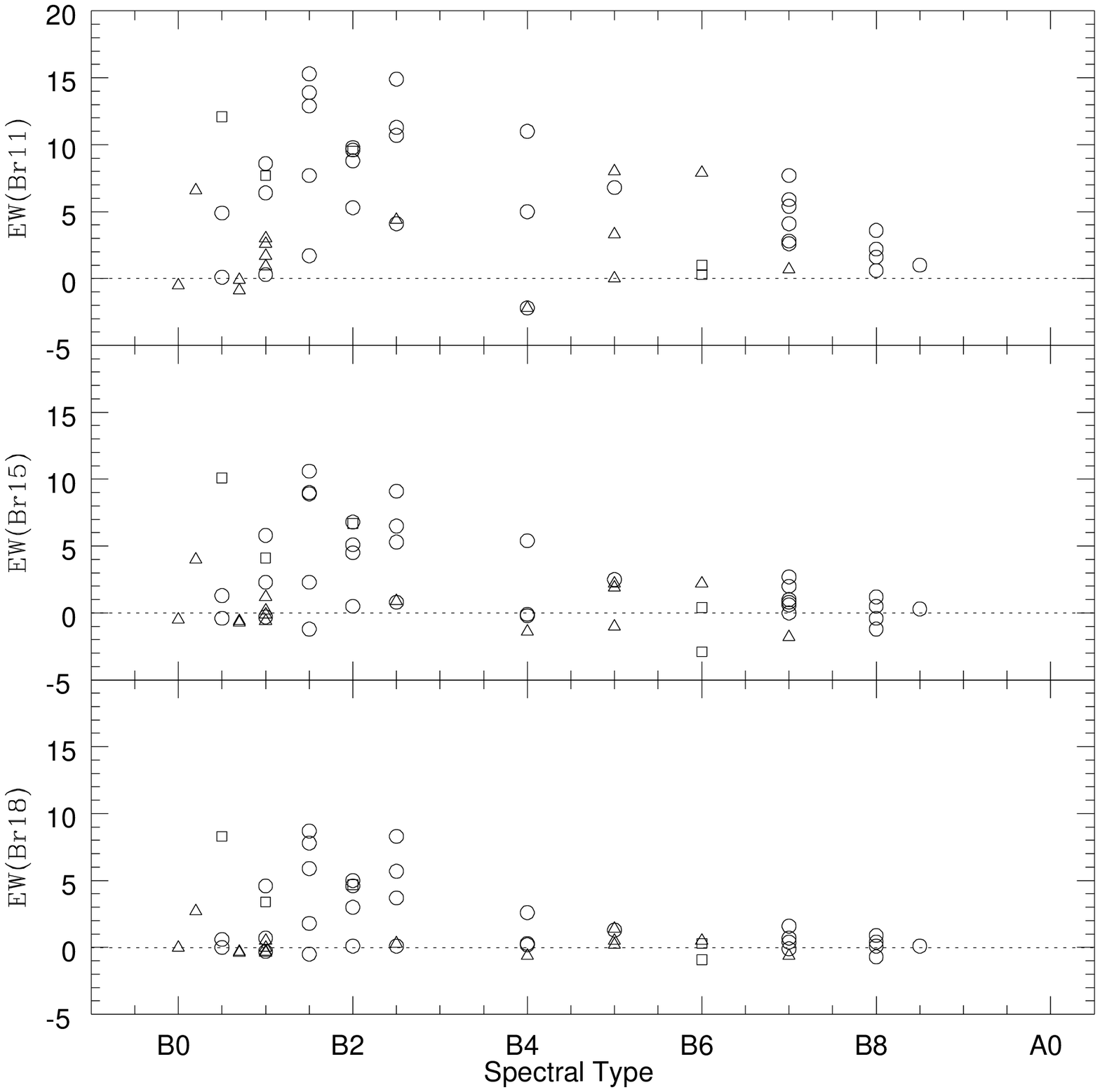}}
\end{picture}
\caption{Equivalent width (EW) of Br-11 (top panel), Br-15 
(middle panel ) and Br-18 (bottom panel) in {\AA} (where positive
EW's indicate emission) against spectral type.  The equivalent widths
have been corrected for the underlying photospheric absorption 
as indicated in the text.  Triangular symbols represent luminosity class III, 
squares luminosity class IV and circles luminosity class V.}
\end{figure}

\def\epsfsize#1#2{0.45#1}
\begin{figure}
\setlength{\unitlength}{1.0in}
\centering
\begin{picture}(3.3,3.3)(0,0)
\put(-0.2,-0.7){\epsfbox[0 0 2 2]{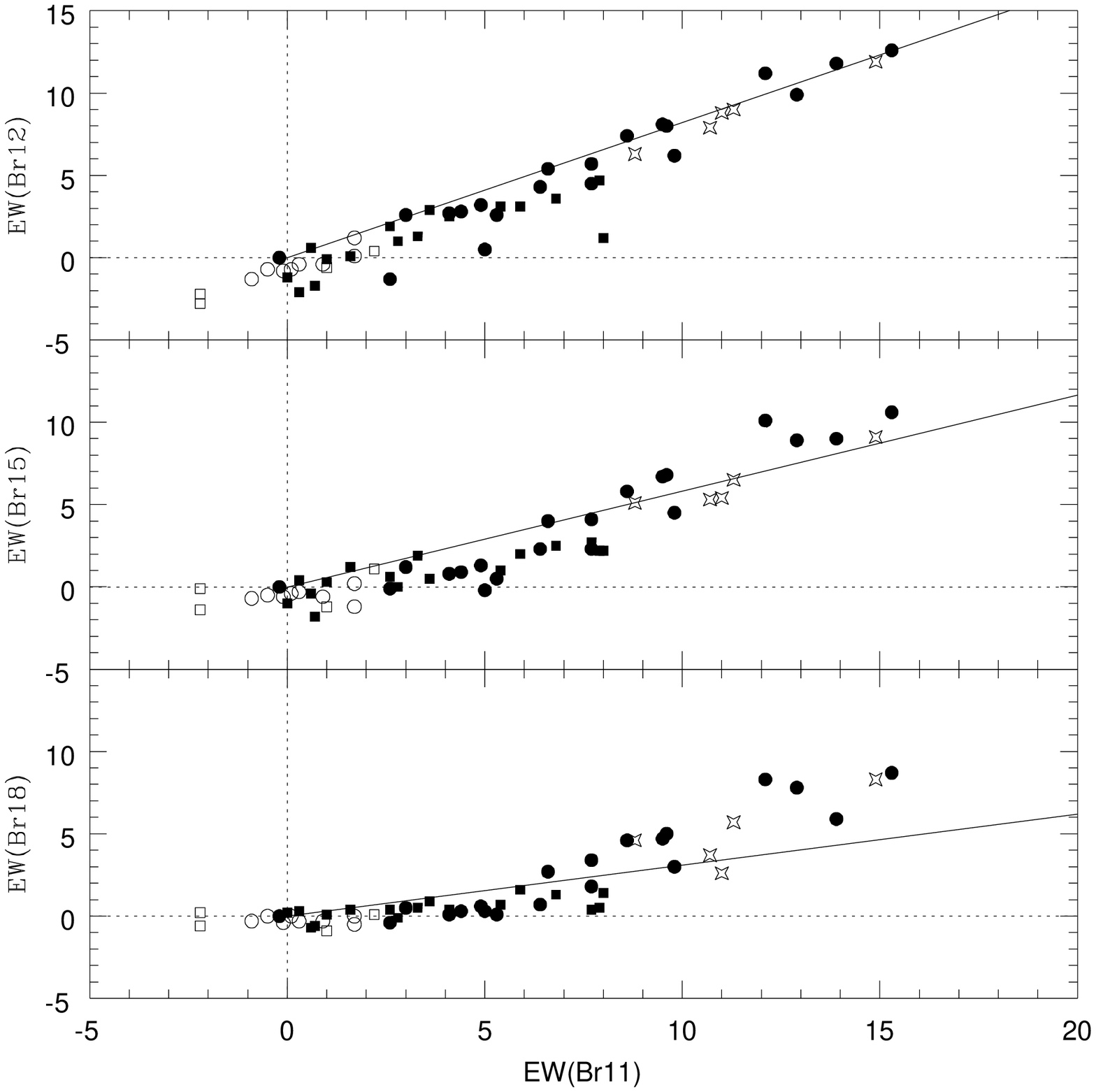}}
\end{picture}
\caption{EW of Br-11 against Br-12 (top panel), Br-15
(middle panel)
and Br-18 (bottom panel). Filled circles represent objects 
from Groups 1, 
circles Group 2, stars Group 3,  empty squares Group 4 and
filled 
squares Group 5 (see Table 1 for group definitions). 
The line indicates the expected line ratio for case B recombination.}
\end{figure}

\def\epsfsize#1#2{0.45#1}
\begin{figure}
\setlength{\unitlength}{1.0in}
\centering
\begin{picture}(3.3,3.3)(0,0)
\put(-0.2,-0.7){\epsfbox[0 0 2 2]{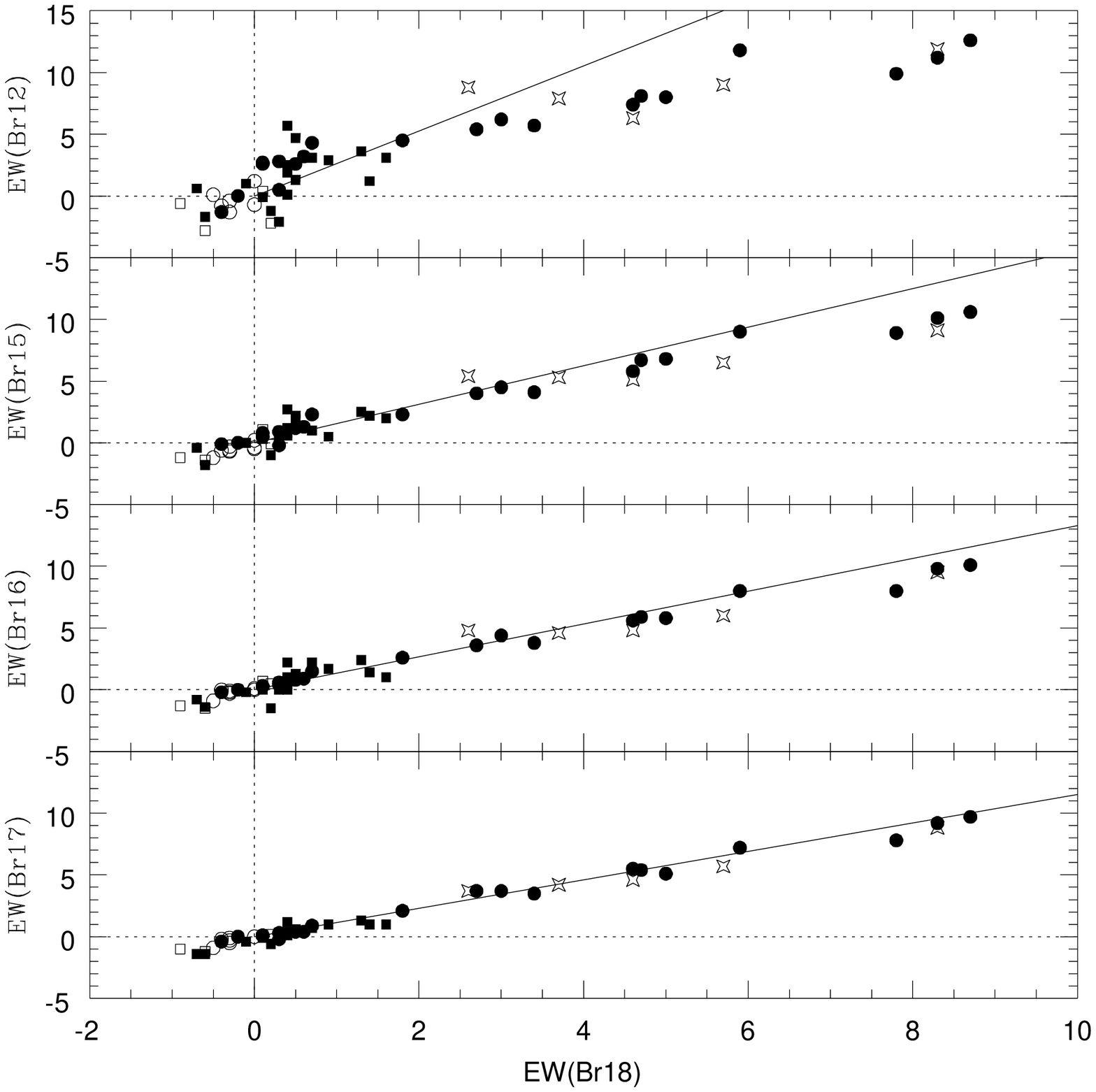}}
\end{picture}
\caption{EW of Br-18 against Br-12, -15, -16 and -17. 
Symbols and line as Fig. 8.}
\end{figure}

\def\epsfsize#1#2{0.45#1}
\begin{figure}
\setlength{\unitlength}{1.0in}
\centering
\begin{picture}(3.3,3.3)(0,0)
\put(-0.2,-0.7){\epsfbox[0 0 2 2]{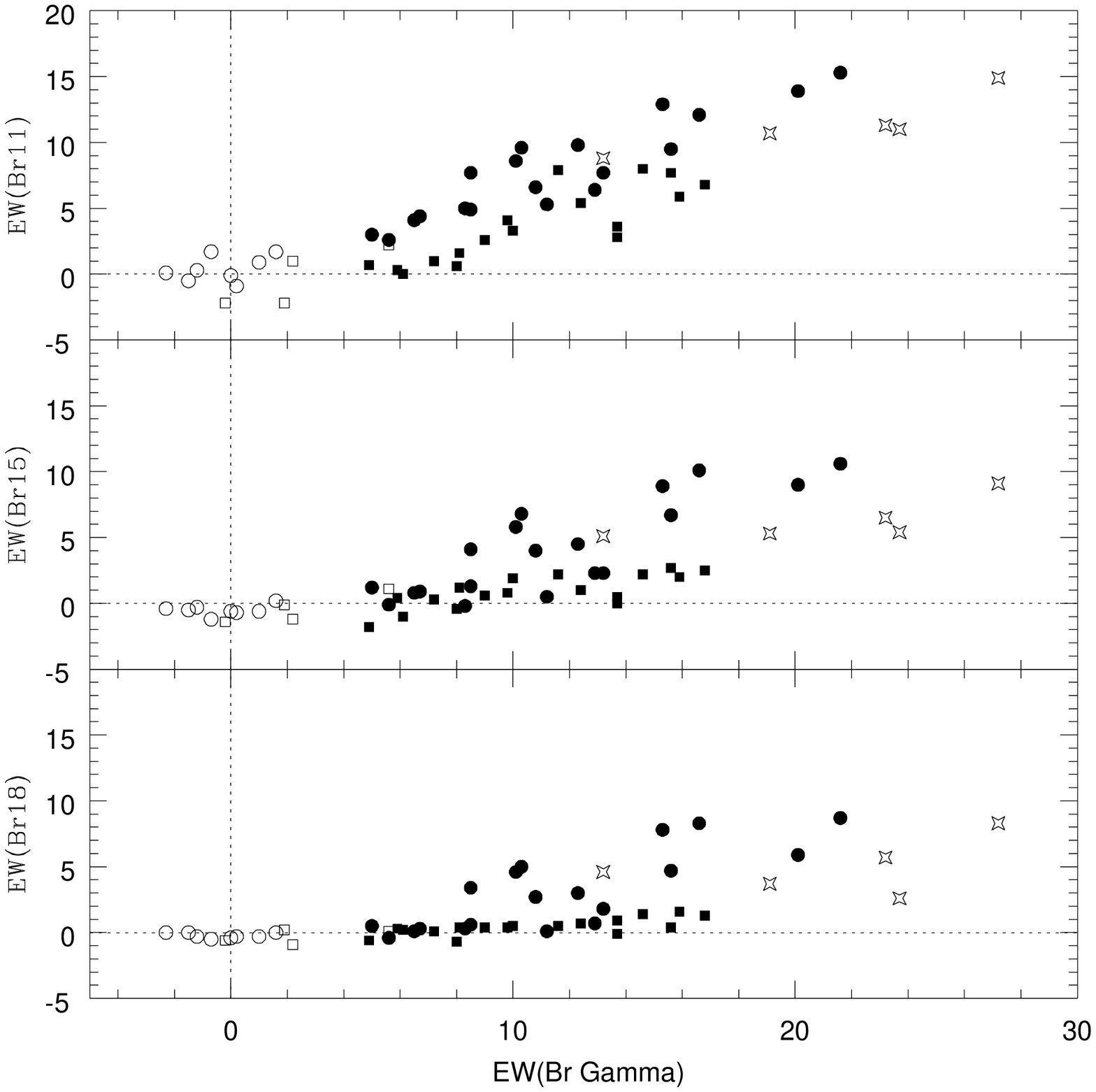}}
\end{picture}
\caption{ Plot of the EW of Br$\gamma$ (data from Paper II)
 against Br-11 (top panel), Br-15 (middle panel)
 and Br-18 (bottom panel; all EW in -{\AA}). Symbols as
 Fig. 8.}
\end{figure}

\subsection{Brackett Line Strengths}

In order to make an accurate comparison of Brackett line strengths
between objects of different spectral type, it is necessary to remove
the effect of the underlying photospheric absorption lines from the spectra.
To carry out this correction we used the equivalent widths for
normal (non-emission) B stars presented for Br-$\gamma$ by 
Hanson et al. (1996) and for Br 11-18 by 
Steele \& Clark(2001).  The correction itself simply consisted of
subtracting the equivalent width derived from a least squares fit
to the appropriate transition for the normal B stars from the emission
line equivalent width.  The discussion and figures in this section
therefore refer to this corrected equivalent width.  We note that
within the small wavelength range encompassed by 
Br-11 to Br-18, these corrected equivalent widths may be divided 
by one another and 
treated as flux ratios without introducing significant error.

We plot the EW of Br-11, 14 \& 18 against spectral type in Fig. 7 (the 
remainder of the lines are not plotted due to reasons of space). As with  
Br$\gamma$ (Paper II) we find that while a linear correlation between spectral 
type and EW is absent, a strong trend in the upper envelope of the
line strengths is present, with lower mean and maximum line strengths
for the later ($\sim$B5--B9) spectral types. 
We find no evidence of systematic differences between the 
emission characteristics  of stars of differing luminosity classes. 

In Fig. 8 we plot the EW of Br-11 against those of Br-12, -15 and -18, 
while in Fig. 9 we plot the EW of Br-18 versus Br-12, -15, -16 and -17.
Note that the empty symbols cluster around EW of zero in both papers, with
a typical deviation of $\sim 1$ {\AA}.  As these are non-emission 
line objects, this indicates that the accuracy of our correction
for the photospheric absorption component is of that order.  We also
note that there appears to be no significant difference in these
graphs between the various spectral groups.

Given that the Br 11-18 lines arise from high levels in the
hydrogen atom they might be expected to be optically thin, and therefore
well represented by case B recombination theory.  In order to
test this, we plot as a straight line 
in Fig.s 8 and 9 the expected case B line ratios
from Storey \& Hummer (1995) for ionized hydrogen.  
The predicted
line ratios are constant to within a few per-cent for temperatures in the
range 7500 - 40000 $K$ and
electron densities in the range $10^{10} - 10^{14}$ cm$^{-3}$, 
and are therefore
should not be sensitive to these conditions within the disk.
From the figures it is apparent that although there is reasonable
agreement between case B theory and observation 
in the line ratios between closely adjacent lines
(e.g. Br-11 and Br-12, or Br-17 and Br-18), for more separated
transitions 
the case B ratios do not 
seem to fit the data well.  This may reflect either
a complex disk temperature and density profile (as discussed
in the following paragraph) or that the optically thin and/or LTE
assumptions for these lines are not valid.  Spectra extending further down the
series (e.g. to Br-22) where the lines are weaker 
would help to resolve this question.

In Fig. 10 we plot the EW of Br$\gamma$ (taken from Paper II)
against that of Br-11, 15 \& 18.  Unlike the higher
transition ratios of Fig. 2 and 3 here there 
are large differences in the flux ratios between the groups, with
a progression in the ratio Br-15/Br$\gamma$ 
from $\sim 0.5$ (Group 1), through 
$\sim 0.2$ (group 3) to $\sim0.0$ (Group 5).
This effect is a real reflection of differences in
line {\em flux} and not just due to continuum differences over the
larger wavelength range for the Br-15/Br$\gamma$ equivalent width ratio.
At first sight it may seem that the stronger circumstellar free-free
excesses of the Group 1 objects could account for this effect, by
reducing the Br$\gamma$ EW for these objects.  However this effect
is small. The two continuum regions
are well approximated by the photometric $H$ and $K$ bands, and
Howells et al. (2001) show that the mean $E(H-K)_{cs}$ only varies 
from $\sim 0.12$ for Group 1 objects to $\sim 0.03$ for Group 5 objects,
insufficient to cause the observed EW ratio changes.  We also note that
this effect is somewhat mitigated by the change in intrinsic
colours between Groups 1 ($(H-K)_o \sim - 0.04$) and 5 ($(H-K)_o \sim
-0.01$) (Koornneef 1983). 
We therefore 
believe that this progression is likely to be
due to systematic changes in the temperature, degree
of ionization, and structure 
within the discs as the temperature and flux of the underlying stars changes.
Marlborough et al. (1997) present simulations of selected near--IR H\,{\sc i}
transitions for both isothermal discs, and discs with a simple radial
temperature gradient, and find that varying the disc temperature does indeed 
lead to changes in the line decrement. However, recent work by Millar \&
Marlborough (1998) shows that these simple models for the disc temperature
are incorrect, and that the equilibrium temperatures for Be star discs are
a complex function of disc radius, particularly in the inner disc regions 
responsible for near--IR recombination line emission.  Further evidence for
this is found in our inability to fit case B line ratios to the
higher transitions as noted previously.
Consequently, we defer
detailed discussion of the systematic variations in the 
Brackett line decrement for a future paper, where 
modeling of the line fluxes will be accomplished for the full
optical--IR dataset.

\def\epsfsize#1#2{0.75#1}
\begin{figure}
\setlength{\unitlength}{1.0in}
\centering
\begin{picture}(3.3,3.0)(0,0)
\put(-0.3,-1.5){\epsfbox[0 0 2 2]{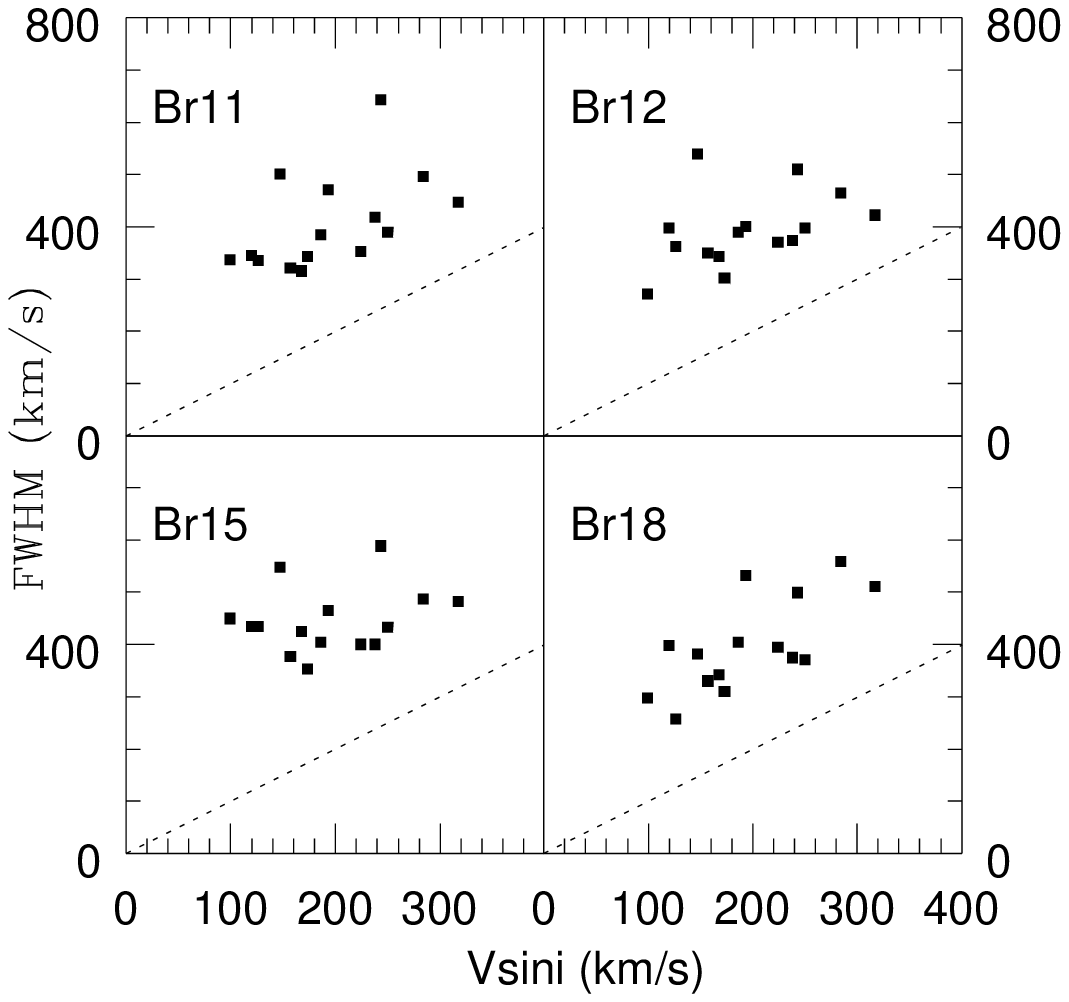}}
\end{picture}
\caption{ Plot of the FWHM (in {\AA}) of Br-11, 12, 15 and 18 
plotted against projected 
rotational velocity. Dotted line indicates FWHM={\em v}sin{\em i} from Paper 
I.}
\end{figure}

\subsection{Line widths and projected rotational velocities}
 
In Paper II we found correlations between the full width half maximum (FWHM) 
of Br$\gamma$ and He\,{\sc i} 2.058$\mu$m lines and the projected rotational 
velocity of the stars, in a similar fashion that those found for H$\alpha$,
$\beta$,  and selected optical Fe\, {\sc ii} lines by Hanuschik (1989). He 
further found that the intrinsically weaker lines such as the higher Balmer 
lines and Fe\,{\sc ii} were systematically broader than H$\alpha$ which he 
attributed to the weaker lines being produced in a smaller region of a 
quasi- Keplerian  disc than the intrinsically 
stronger H$\alpha$ line, where the rotational velocity would be greater, and 
thus the line broader. Simulations by Marlborough et al. (1997)
also demonstrate the same effect, as do the longer wavelength 
infrared spectra of $\gamma$-Cas presented by Hony et al. (2000).

In Fig. 11 we plot the full width half maximum (FWHM) of Br-11, 12, 15 
and 18 for stars of Groups  1 and 3 against 
their projected rotational velocities from Paper 1 (stars from Group 5 were 
excluded due to the bias introduced into the FWHM by the underlying 
photospheric feature).  
Weak correlations were found between the FWHM and the 
projected rotational velocities  by applying Spearman's Rank Correlation to 
the datasets, and the results of least square fits to the 
data are presented in Table 7.  
It is interesting to note that the minimum
measured FWHM is around 300 km/s (considerably 
larger than the instrumental resolution of $\sim 100$ km/s). This means
that for the relatively slow ($v \sin i \sim 100-200$ km/s) 
rotators, the FWHM for these lines is often more than twice 
the $v \sin i$ of the underlying star, possibly indicating more rapid rotation
at the inner edge of the disk than at the surface of the star.
We also note three Group 1 (BD +47 3985, BD +58 2320 and BD +1 1005) 
and one Group 3 (BD +42 4538) appear to have very large Br-11 FWHM of greater
than around 500 km/s.  However these objects also show 
double peaks in their line 
profiles, making the derivation (and interpretation) of FWHM
uncertain. 
 
In Fig. 12 we plot the FWHM of Br-18, 15 \& 12 vs Br-11.  It can be seen that 
there is no trend to 
broader line profiles with higher Brackett series transitions as was 
suggested by the results of Hony et al. (2000). 
We attribute this result primarily 
to the narrow wavelength (and hence transition) range of our spectra
compared to the work of Hony et al. 
(2000) whose spectra extend to the limit of each series.

\begin{table}
\begin{center}
\caption{Summary of best fits between the FWHM of Br11--18 and the 
{\em v}sin{\em i} of the individual star. The best fit from Br$\gamma$, taken 
from Paper II is also shown for comparison.}
\begin{tabular}{cc}
\hline
Transition & Best Fit \\
\hline
Br18 & 1.132{\em v}sin{\em i} +181{$kms^{-1}$} \\
Br17 & 1.145{\em v}sin{\em i} +203{$kms^{-1}$} \\
Br16 & 0.868{\em v}sin{\em i} +280{$kms^{-1}$} \\
Br15 & 0.337{\em v}sin{\em i} +377{$kms^{-1}$} \\
Br14 & 0.607{\em v}sin{\em i} +304{$kms^{-1}$} \\
Br13 & 0.541{\em v}sin{\em i} +301{$kms^{-1}$} \\
Br12 & 0.553{\em v}sin{\em i} +283{$kms^{-1}$} \\
Br11 & 0.803{\em v}sin{\em i} +245{$kms^{-1}$} \\
\hline
Br$\gamma$ & 0.759{\em v}sin{\em i} +149{$kms^{-1}$} \\
\hline
\end{tabular}
\end{center}
\end{table}

\section{Conclusions}

\def\epsfsize#1#2{0.45#1}
\begin{figure}
\setlength{\unitlength}{1.0in}
\centering
\begin{picture}(3.3,3.3)(0,0)
\put(-0.1,-0.8){\epsfbox[0 0 2 2]{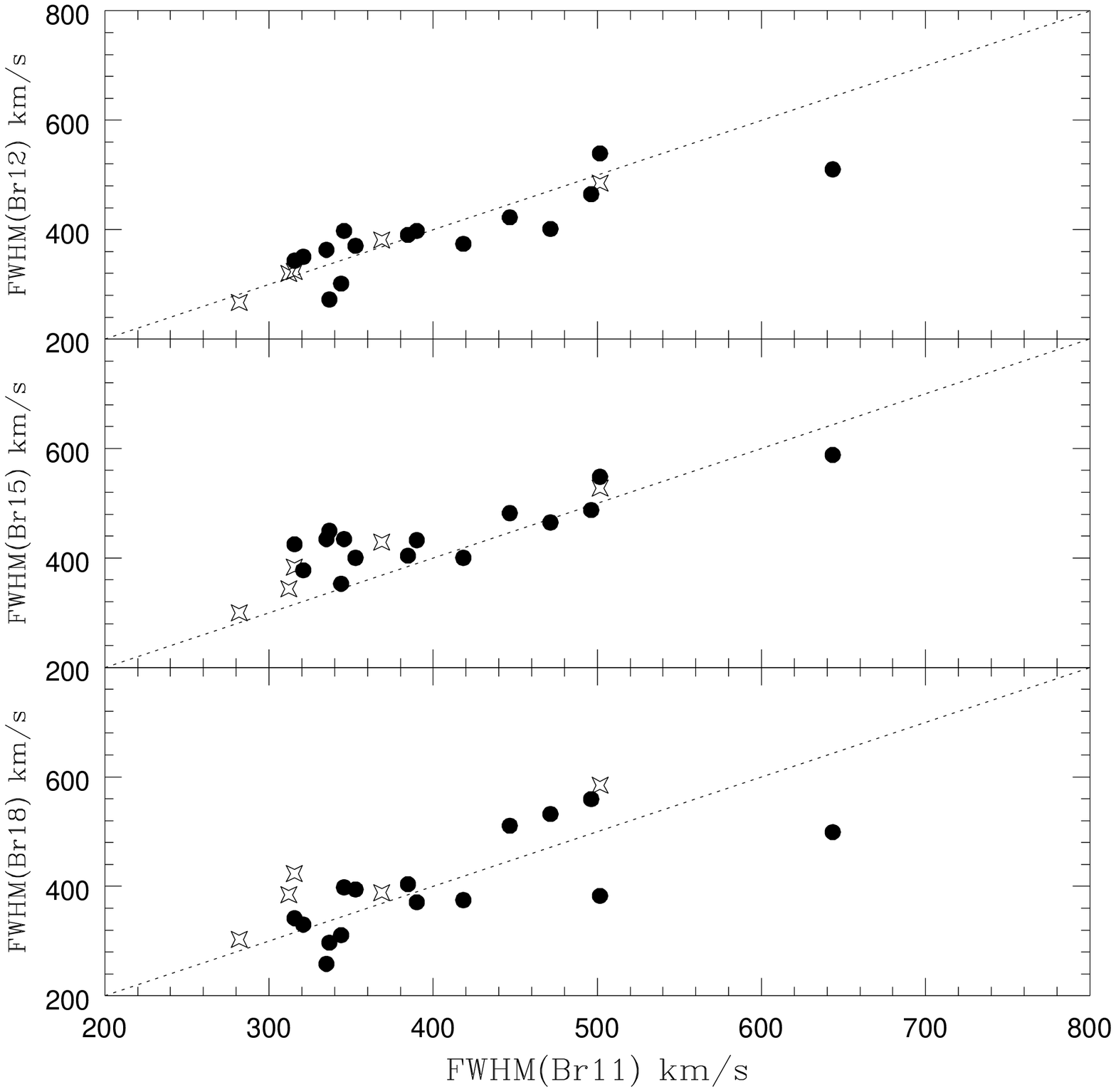}}
\end{picture}
\caption{FWHM of Br-18, 15 \& 12 vs Br-11 for the objects showing
emission features.  No systematic difference between the FWHM of Br-11, and 
those of the higher transitions is apparent. The larger scatter of points
in the lower panel is likely to be due to the weakness of the Br-18 line,
and the consequent difficulties in obtaining an accurate measurement of the 
FWHM.  Symbols as Fig. 8.}
\end{figure}

We have analysed the H band (1.53$\mu$m--1.69$\mu$m) spectra of 61 Be stars 
and found emission from  H\,{\sc i} Br11--18 and Fe\,{\sc ii} 1.576 \& 
1.698$\mu$m. A further emission feature at 1.679$\mu$m is
present in a subset of 14 stars; this is also likely  to be Fe\,{\sc
ii} emission.  The Br 11--18 line ratios of non adjacent lines 
are not well fit by the
case B assumption, and strong systematic trends in both line strength 
and the ratio of the higher Brackett series strengths to Br-$\gamma$
with spectral type (but not luminosity class) are observed. 
This is likely to be due to systematic changes in the temperature and degree
of ionization between the discs of stars of early (B0--B4) and late (B5--B9)
spectral type.
We find that analysis of  H band spectra alone
only allow the classification of stars into ``early'' (B0e--B4e) or late 
(B5e--B9e) types; no determination of the luminosity class of the object can 
be made. This is due to the lack of any uncontaminated photospheric features
in this region, the lack of emission features encompassing a wide range of 
excitation temperatures and the similarity of Brackett line strengths and 
decrements for the early type stars. 
As with Br$\gamma$, we find weak 
correlations between the FWHM of Br11-18 and the 
projected rotational velocity of the underlying stars. We 
find no systematic trend in FWHM through the Brackett 
series down to Br-18. 

\section{Acknowledgements}
UKIRT is
operated by the Joint Astronomy Centre, Hawaii for the UK PPARC.
We thank the support astronomers and staff
of UKIRT for their invaluable assistance at the telescope.  
Data reduction and analysis for this paper was carried out using
the Liverpool JMU and Sussex University Starlink Nodes. 
JSC wishes to acknowledge a PPARC research award.

\end{document}